\documentstyle[aps,prl,twocolumn,psfig]{revtex}
\begin{document}
\twocolumn[\hsize\textwidth\columnwidth\hsize\csname@twocolumnfalse\endcsname
\title{\bf  Coulomb blockade in low mobility nanometer size Si:MOSFETs}
\author{ M. SANQUER,  M. SPECHT, L. GHENIM , \\ S. DELEONIBUS* and G. GUEGAN*}
\address{ CEA-DSM-DRFMC-SPSMS, \\ and *CEA-DTA-LETI-DMEL CEA-Grenoble \\  38054 Grenoble Cedex, France.}
\maketitle
\begin{abstract}
We  investigate coherent transport in Si:MOSFETs with nominal gate lengths 50 to 100nm and 
various widths at
very low temperature. Independent of the geometry, localized states appear 
when $G \simeq e^{2}/h $ and transport is dominated by resonant tunnelling 
through a single quantum dot formed by an impurity potential. We find that the typical size of 
the relevant impurity quantum dot is comparable to the channel length and that the periodicity of the observed Coulomb blockade oscillations 
is roughly inversely proportional to the channel length. The spectrum of resonances and the nonlinear I-V curves allow to measure the 
charging energy and the mean level energy spacing for 
electrons in the localized state. Furthermore, we find that in the dielectric 
regime the variance $var(lng)$ of the logarithmic conductance $lng$ is proportional to its average value $<\!lng\!>$ consistent with one-electron 
scaling models. \pacs{73.23.-b, 73.23.Hk, 72.15.Ra}
\end{abstract}]

\par After the pioneering work of Scott-Thomas et al \cite{scott-thomas}, Coulomb blockade in quantum dots formed by an impurity potential has been studied in quasi 1D wires
or point-contact geometries \cite{degraaf,nicholls}. In comparison with lithographically defined lateral quantum dots, impurity quantum dots (IQD) contain typically fewer electrons. Downscaling the size of the IQD allows to operate a silicon based single electron quantum dot transistor even at room temperature \cite{zhuang}\cite{ishikuro}.
In the opposite case of wires, i.e. in disordered thin and wide insulating barriers, no Coulomb blockade oscillations have been reported up to now.
Resonant tunneling through single ionized donor potentials is responsible for electron transport in disordered thin insulating barriers. This has been studied in the deeply insulating regime of large thin barriers formed by depleting electrostatically a semiconductor under a gate \cite{kuznetsov}, or in thin amorphous silicon tunnel barriers \cite{beasley}.
Interaction between distant impurity states in the channel have been revealed by peculiarities of the nonlinear transport. However, a single ionized donor potential cannot accomodate many electrons without becoming screened.

\par We report for the first time Coulomb oscillations in  very short  MOSFETs with source drain distance $d_{SD}$ less than $0.05 \mu$m and width much larger than $d_{SD}$. Contrarily to quasi 1D wires, where the size of the IQD is somewhat arbitrary, we will show that the diameter of the IQD is comparable to the source drain distance and that on resonance the conductance g in quantum units $e^{2}/h$ is close to one.  Furthermore we demonstrate that the fluctuations of the conductance characterized 
by $var(lng)$ are proportional to the mean value $<\!lng\!>$, rather independent of geometry. Such an observation is consistent with one-parameter noninteracting scaling models of the metal-insulator transition. Our 
experimental findings suggest that interactions do not destroy this one-parameter description.

\par The devices are MOSFETs  on the (100) surface
of silicon doped to a level of $7 \ \ 10^{12}Borons/cm^{-2}$ for the 100nm series and of $3 \ \ 10^{13}Borons/cm^{-2}$ for the 50nm series.
The gate oxyde thickness 
is only $d_{SiO_{2}}=3.8 nm$ for the 100nm series and $d_{SiO_{2}}=2.4 nm$ for the 50nm series. This screens strongly Coulomb interactions in the inversion layer. Source and drain consist of highly  ion implanted 
regions ($ 10^{15}As/cm^{-2}$).   The polysilicon gate has a length 
of 100nm or 50nm and the transverse dimension varies between 300 nm and 25 $\mu m$. The 
effective channel length $d_{SD}$ between source and drain is somewhat smaller than the geometrical
gate length due to extension regions ($2 \ \ 10^{14} As/cm^{-2}$). $d_{SD}$ is estimated to be 
of order 25 nm and 75 nm in the two cases considered here. 
The room temperature mobility is  $242 \,cm^{2}s^{-1}V^{-1} $ for the 100nm series and $150 \, cm^{2}s^{-1}V^{-1} $ for the 50nm series.
The source-drain
current $I_{SD}$ is measured for a source-drain voltage $V_{SD} = 10 \mu V$ as 
a function of the gate voltage using a standard low-frequency lock-in 
technique. The $I_{SD}-V_{SD}$ characteristics is linear for this value of $V_{SD}$  even 
at the lowest temperature, independent of $V_{g}$. The sample is inside a copper box 
thermally anchored to the mixing chamber of a dilution refrigerator. About 2 meters 
of Thermocoax Philips \cite{zorin} on each side of the sample provide the contact to the 
lock-in amplifier.  

\par
\begin{figure}
\psfig{figure=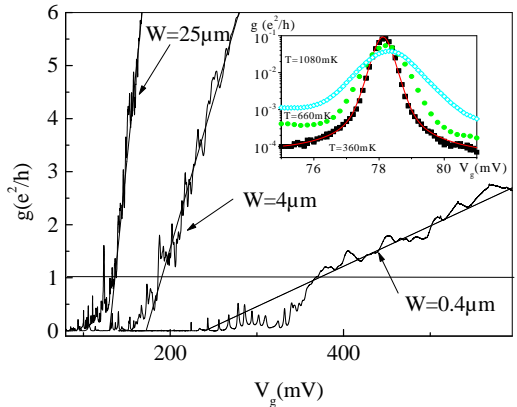,width=87mm}
\caption{ Source-drain conductance versus gate voltage of samples with $L = 100 nm$ and various widths W  at
T=35 mK. The horizontal line corresponds to $G = e^{2}/h$ where independent of geometry diffusive transport sets in. A linear dependence of $g$ with $V_{g}$ is expected if one supposes a constant mobility, a constant density of states and a constant gate capacitance.  Inset: Source-drain conductance versus gate voltage at various 
temperatures for a resonance in the dielectric regime of 
the W=4$\mu$m sample. The solid lines is the fit associated to thermally broadened resonant tunnelling (see text) \protect\cite{fitcomment}. The intrinsic linewidth 
infered from the fit is $\Gamma_{e} / k_{B} = 60 mK$.
}
\label{fig1san}
\end{figure}

 Figure \ref{fig1san} shows the source-drain conductance in quantum units 
versus $V_{g}$ in three  samples of gate length $L=100 nm $ differing only by the width. At a
temperature of $T=35$mK, we observe reproducible conductance fluctuations as a function of gate voltage which  persist up 
to $T=20$K at small  $V_{g}$. Depending on the 
value of the conductance these fluctuations evolve differently with temperature:
If $G > e^{2}/h $ the fluctuations are 
gaussian around their mean value and poorly sensitive to temperature below  T=4.2K. This characterizes universal conductance 
fluctuations in the diffusive regime not further considered here.
If $G < e^{2}/h $, fluctuations evolve into sharp resonances at very low 
temperature (dielectric regime). 
In the diffusive regime, the linear increase of the conductance with the gate voltage  permits to evaluate the mobility $\mu$: $ G_{square} = n_{s}e \mu = C_{g} V_{g} \mu \propto V_{g}$ if $\mu$ and $C_{g}$ do not depend on $ V_{g}$. We find $\mu = 25 \ \  cm^{2}V^{-1}s^{-1}$ for  $ 0.135V \le V_{g} \le 0.2V$ in the $W=25 \mu m $ sample, $\mu = 53 \ \ cm^{2}V^{-1}s^{-1}$ for  $ 0.2V \le V_{g} \le 0.3V$ in the $W=4 \mu m $ sample and $\mu = 73 \ \ cm^{2}V^{-1}s^{-1}$ for  $ 0.375V \le V_{g} \le 0.6V$ in the $W=0.4 \mu m $ sample (at low temperature). These mobilities are weak reflecting a strong short range disorder and a mean free path of order $ 4 nm$.  The measured capacitance to gate coincides with the theoretical estimations: $C_{g}/S ={\epsilon_{0} \epsilon_{SiO_{2}} \over d_{SiO_{2}}} 
=  10^{-14} F/ \mu m^{-2} $ (for $d_{SiO_{2}}=3.8 nm$). A variation of $\delta V_{g}= 1V$ induces
 $\delta n_{s}= 6.25 \ \ 10^{12} cm^{-2}$ for the 100nm series ($\delta n_{s}= 1.0 \ \ 10^{13} cm^{-2}$ for the 50nm series) \cite{threshold}.

\par Typical conductance resonances in the dielectric regime are well fitted over 3 orders of 
magnitude (see inset of Fig.1) 
by the standard relation for the thermally broadened resonant tunneling regime:
\begin{eqnarray}
G( V_{g},T)  & = & {e^{2}\over h} A \int\limits_{-\infty}^{\infty} L(V)\times {\partial 
f( e \alpha  (V-V_{g}), T) \over \partial V} dV \nonumber \\ 
L(V) & = & {\Gamma_{e}^{2}
\over {e^2 \alpha ^2 (V_0 - V)}^{2} + \Gamma_{e}^{2}} \nonumber
\end{eqnarray}
with $f( x, T)= ( 1 + exp({ x/k_{B}T}))^{-1}$,
 $\alpha = 0.252$ and $A= 0.8 \pm 0.2$.
 $\Gamma_{e} / k_{B} = 60 mK$ is the intrinsic linewidth and $V_{0}$ is the gate voltage at the resonance.
  $ \delta E_{F} = e \alpha \delta V_{g}$ is the 
variation of the Fermi energy. $T=360 mK \pm 15mK$ is the effective electron
temperature. $A \simeq 1$ means that the localized state is equally connected to the source and the drain. If  not, the resonant 
conductance would be  exponentially smaller\cite{azbel}.
 From  $\alpha =  C_{g}/ e^{2} g_{2D} $, we estimate an approximately constant density of states in the channel $g_{2D} \simeq   0.25 \ \ 10 ^{14}cm^{-2}eV^{-1} $, 
which is reduced  by a factor 4 compared to the metallic 2D density of states. This is due to Lifshitz tails induced by disorder at the bottom of the conduction band.

\par
\begin{figure}
\psfig{figure=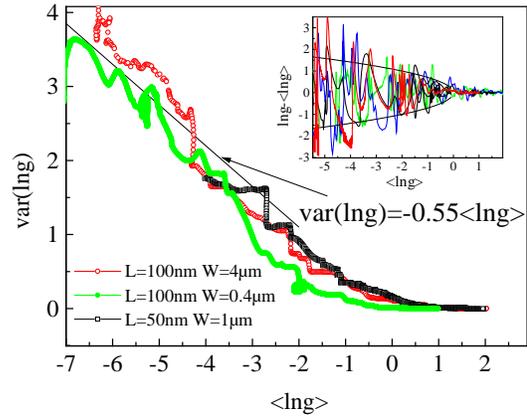,width=87mm}
\caption{ The variance $var(lng)$ of the logarithmic conductance $lng$ versus its average  
$<\!lng\!>$ for 4 samples with different geometries. For $<\! lng\! > \le 0$, $lng$ fluctuates strongly roughly corresponding to $var(lng) \simeq -0.55<\!lng\!>$ over orders of magnitude in the insulating regime.
Inset: Fluctuations of the logarithmic conductance versus $<\!lng\!>$ of four samples with various 
geometries. The solid lines correspond to $\pm \sqrt{-0.55<\!lng\!>}$. For $<\!ln(g)\!> \ge 0$, the log-normal fluctuations of the dielectric regime disappear independent of geometry.}
\label{fig2san}
\end{figure}

 The central result in Fig. \ref{fig1san} is
that the onset of diffusive transport  is essentially 
independent of the width of the sample. It always occurs when the total conductance $G \simeq e^{2}/h $. This is also illustrated in a scaled representation of the fluctuations 
with respect to the average conductance $<\!lng\!>$ in the inset 
of Fig.~\ref{fig2san} : Independent of geometry the 
log-normal fluctuations of lng disappear  when  $<\!lng\!>\, \geq 0 $. 

\par The explanation of this striking result lies in the very broad distribution of conductances in the dielectric regime. Theories of log-normal distributions of conductances \cite{raikh} 
indicate that an anomalously large conductance, a rare event, dominates  the total conductance.  This is 
only valid at very low temperature, when parallel thermally activated conducting channels 
are exponentially small.
According to the  scaling theory of localization, at the metal-insulator transition, this event is the last conducting channel with conductance $G \simeq e^{2}/h$ \cite{imry}. 
In  samples consisting of many squares  in parallel and since the conductance distribution is log-normal, this last conducting channel carries almost the total current.

\par Deep in the insulating regime, we find that the conductance fluctuations obey 
roughly to: $var(lng) \simeq -0.55<\!lng\!>$, independent of geometry for widths up to $W=4 \mu m$ \cite{wilfrid} (see Fig. \ref{fig2san}). 
Here $var(lng) = <\![lng- <\!lng\!>]^{2}\!>_{V_{g}}$ is the variance of the average conductance $<\!lng\!>$ where the average is taken over ${V_{g}}$ . 
The proportionality between the variance and the mean of $lng$ is expected in disordered insulators at zero temperature and described by non-interacting scaling models\cite{Pichard}. However, since interactions are crucial to explain the 
details of the fluctuations as shown below, our experiment suggests that a one-parameter description of the 
metal-insulator transition (at least in the strongly insulating phase) holds 
even in the presence of interactions. 

\begin{figure}
\psfig{figure=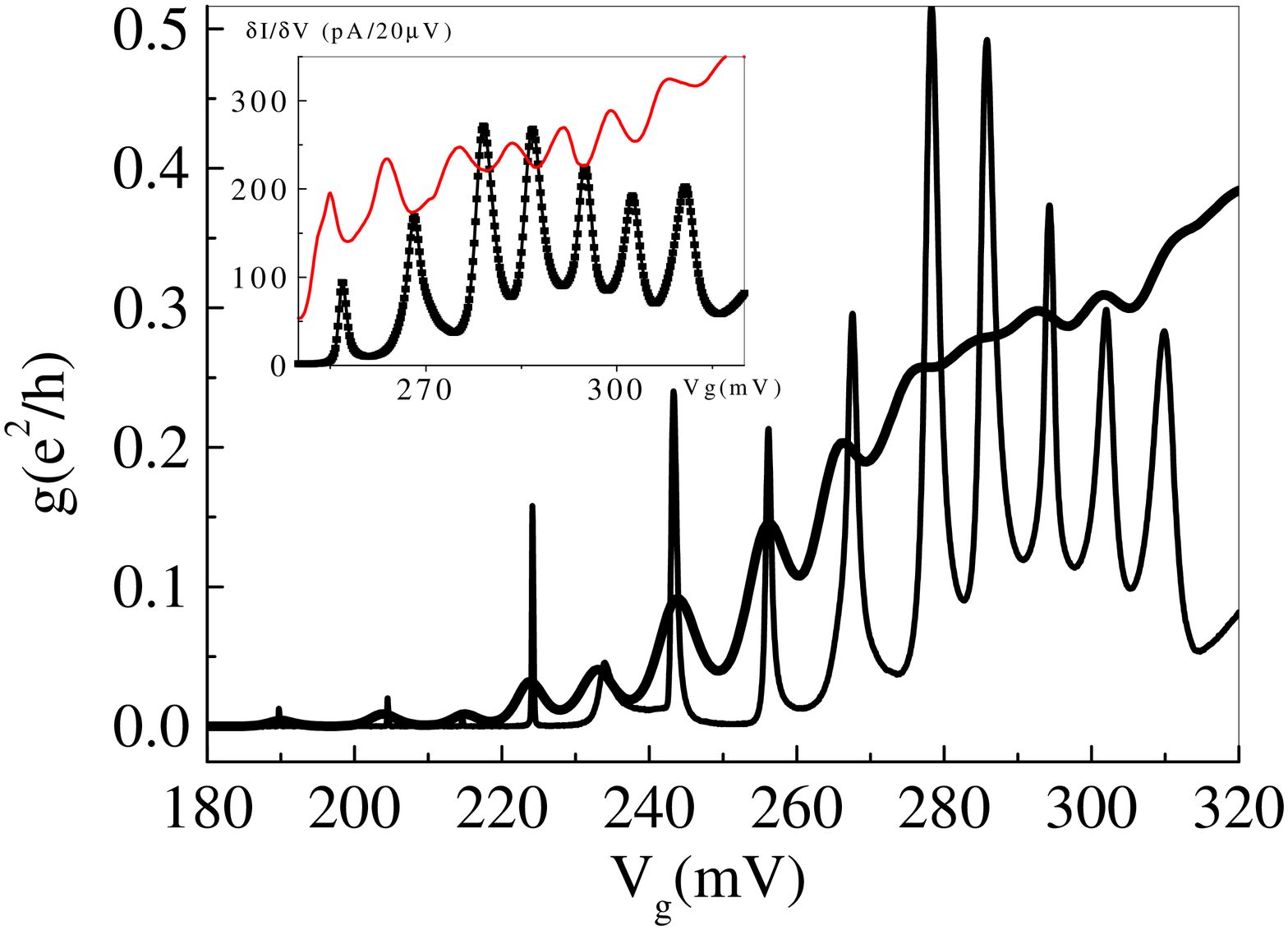,width=87mm}
\psfig{figure=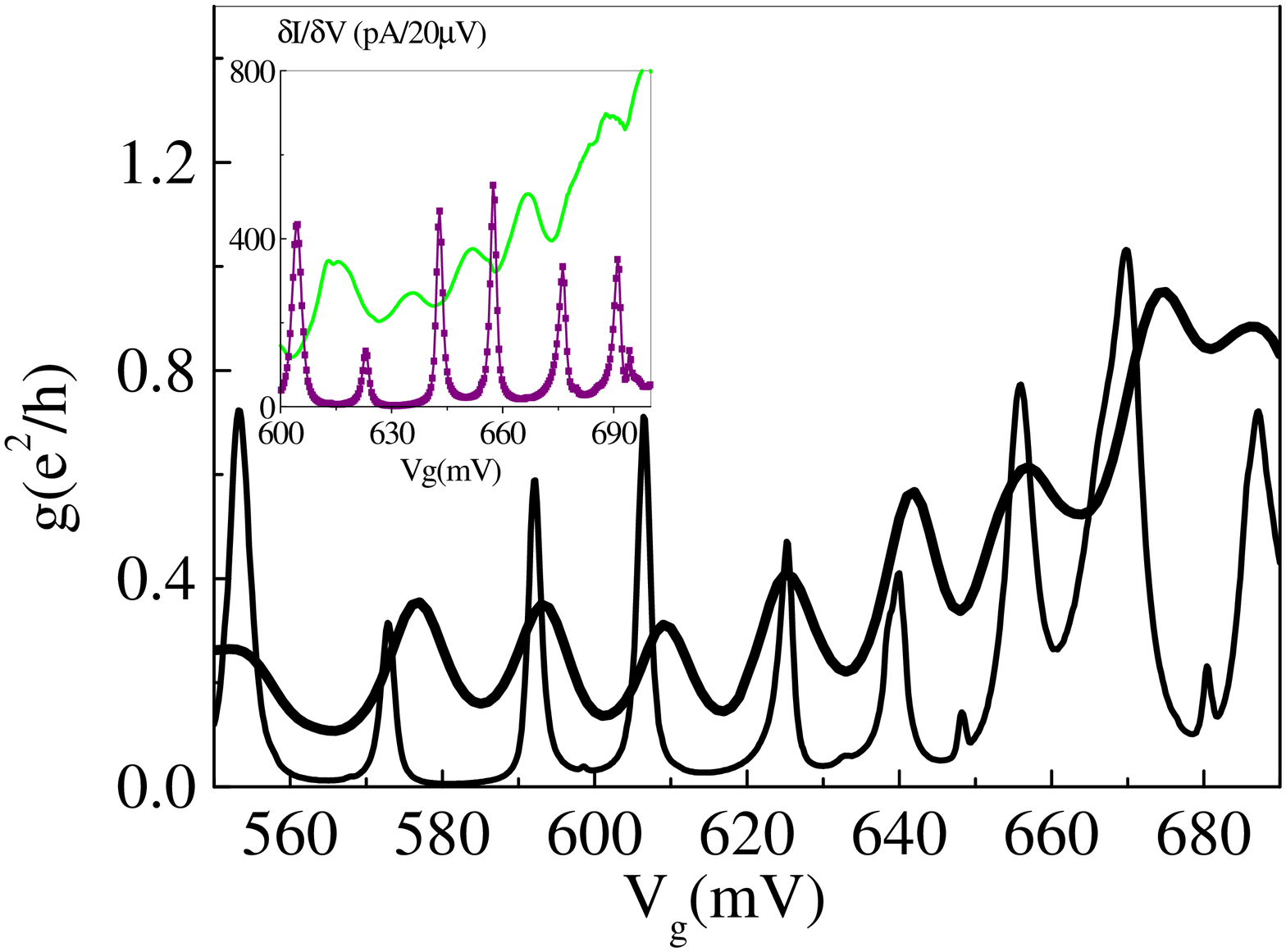,width=87mm}
\caption{ Periodic oscillations of the conductance versus gate voltage in two different samples (TOP: $d_{SD}=75nm$ W=400nm. BOTTOM: $d_{SD}=25nm$, W=1000nm). Thick lines are at T=4.2K and thin lines at T=35mK. The oscillations are periodic at T=4K but not strictly periodic at lower temperature. Top inset: differential conductance oscillations for $V_{SD}=0$ and $V_{SD}=-2mV$ in the 100nm sample at T=35mK. 
 Bottom inset: differential conductance oscillations for $V_{SD}=0$ and $V_{SD}=-3.5mV$ in 
the 50nm sample at T=35mK. Note the alternance of conductance maxima, predicted in the classical Coulomb blockade model.
}
\label{fig3san}
\end{figure}

\par Fig. \ref{fig3san} shows Coulomb oscillations in our shortest samples with effective source drain distances $d_{SD}$ of 
25nm and 75nm. In both samples each conductance peak at $T=4.2K$ corresponds to one peak at the lowest 
temperature. No additional peaks appear upon cooling and we only observe small shifts in gate voltage. This 
is consistent with {\it single} dot resonances and rules out  multiple dot tunnelling  as previously reported
in quasi-1D wires\cite{degraaf,ruzin}. 
Actually, several dots acting in parallel explain qualitatively 
the complicated pattern observed in samples with widths larger than typically $1\mu m$. 

\par We first consider the series of resonances at relatively high gate voltage when the dot is filled 
with several electrons. Taking a simple Coulomb blockade model periodic conductance resonances occur if the temperature is lower than 
the charging energy $E_{C}\gg k_{B}T$. At zero source-drain voltage $V_{SD}$ and $V_{g}$ such that 
$C_{g}V_{g} = Ne +e/2$ where $N$ is the number of electrons in the dot, charge is transfered; 
If on the other hand $V_{SD}=e/C$, resonances occur at $V_{g}$ such that 
$C_{g}V_{g} = Ne$. This is demonstrated in the insets of Fig. 3. We suppose here that the capacitances 
to source and drain $C_{S}= C_{D}=C/2$ are equal. 
Within this classical model, we find 
$  C = 8 \ \ 10^{-17} F$ and $  C_{g} = 2 \ \ 10^{-17} F$ for the 100 nm sample,
and $  C = 4.6 \ \ 10^{-17} F$ and $  C_{g} = 1 \ \ 10^{-17} F$ for the 50 nm sample.  The charging energy is $e^{2} / C \simeq 2meV$ and $3.5meV$ respectively for the 100nm and the 50nm sample.
 In the simplest model, $\alpha =\delta E_{F}/  e \delta V_{g} =  C_{g} / (C_{g} + C) \simeq 0.2$, in agreement with the value  deduced from the temperature dependence of the resonances (see inset of Fig. ~\ref{fig1san} ).

\begin{figure}
\psfig{figure=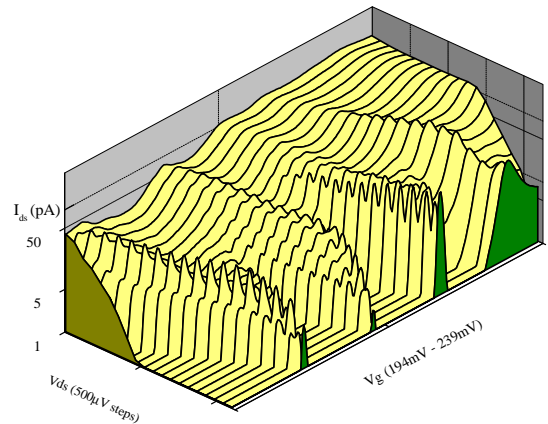,width=87mm}
\caption{ Source-drain current ( AC excitation voltage: $20 \mu V$) versus gate voltage 
at various DC-bias $V_{SD}$ ( 0-9.5mV by steps 0.5mV) for the 
sample (W=0.4$\mu$m, L=0.1$\mu$m) at T=35mK. Several excited states appear at finite  $V_{SD}$.}
\label{fig4san}
\end{figure}

\par With decreasing number of electrons 
 on the dot, the mean level spacing $\Delta$ increases and the barrier resistance 
between source and drain and the impurity quantum dot increases. 
 If the temperature is below the mean level spacing we can estimate $\Delta$ by the 
 $I_{SD}-V_{SD}$ non-linearities revealing the excited states of the dot. In our samples this is 
shown in Figure ~\ref{fig4san}. Upon rising $V_{SD}$, the differential conductance $ \delta I_{SD} / \delta V_{SD}$ exhibits a peak each time an excited state $E_{i}$ enters or exits  the energy window 
$\mu_{S}+ \alpha e V_{g} > E_{i} >  \alpha e V_{g}+\mu_{D}$.
 Up to 3 excited states ( 6 peaks in $ \delta I_{SD} / \delta V_{SD}$) are distinguishable for  resonances at low gate voltage for $V_{SD} \simeq 3 meV$, such that $\Delta \simeq 1meV $. At higher gate voltage $\Delta$ becomes smaller and  the contrast between excited states is washed out (classical Coulomb blockade regime).
Since $\Delta$ is smaller but non negligible compared to $E_{C}$, the periodicity 
 of the Coulomb oscillations is not strictly obeyed.

\par The values for the mean spacing and charging energy are typical for all the measured samples. $\Delta \simeq 1 meV$ corresponds to what is expected for single electron levels in a 2D box of $65nm \times 65nm$ taking into account the reduced density of states in the Lifshitz tail. Alternatively it corresponds also to a parabolic potential confinement of radius 50 nm, comparable to the source-drain distance, and heigth 10 meV, comparable to the Fermi energy at the MIT. The capacitance to gate $ C_{g} = 2 \ \ 10^{-17} F$ (L=100nm) (resp.  $ C_{g} = 1 \ \ 10^{-17} F$ for L=50nm)  corresponds to a 2D dot of size $45nm \times 45nm$ (resp. $25nm \times 25nm$).

\par A main feature of our samples is the very thin gate oxyde which makes the gate very efficient to screen the Coulomb interaction in the IQD. For instance, a simple estimation of the repulsion 
between two {\it bare } electrons at a mean distance of $r= 20nm$ in the 
channel gives $ V= {1 \over 4 \pi \epsilon_{0}} 2 \epsilon_{Si} {d^{2} \over \epsilon_{SiO_2}^{2}}{e^{2} \over r^{3}} \simeq 3.9 meV$
where $\epsilon_{Si}=11.4$ and $\epsilon_{SiO_2} \simeq 4$ are the relative dielectric constants
of $Si$ and $SiO_2$. Since the oxyde thickness ($d_{SiO_2}=3.8nm$) is much smaller than
the distance  between  two electrons, the Coulomb interaction is a dipole interaction.

\par The various above estimations confirm that the sequence of resonances reflects the interaction between electrons sharing the same (barely) localized site whose extension is comparable to the source-drain distance independent of the width of the channel or the dopand concentration. Furthermore, 
we have demonstrated in this way that in order to observe Coulomb blockade in nanostructures it is not necessary to constrict the current through a quasi 1D segment such as point contacts or wires. The domination 
of a rare event in a short 2DEG intrinsically favours a channel through a single IQD in not too wide geometries. In this way one has a better estimate of the relevant length and energyscales of the dot 
than in quasi 1D geometries. Also, downscaling the source-drain distance $d_{SD}$ from 75nm to 25nm increases the Coulomb energy, as expected when the dot size of the charge transmitting IQD decreases with channel length.
 
\par In summary we have studied standard Si:MOSFETs of gate length $L=50nm$ and  $L=100nm$ and various 
widths without any intentional confinement.  Close to the MIT, characterized 
by  $G \simeq e^{2}/h $ at low temperature,  conductance resonances
as a function of the gate voltage are due to tunneling through a single disordered quantum dot, 
whose extension is comparable to the source-drain distance and which accomodates 
several electrons. In contrast to many previous studies on quasi-1d wires, our geometry allows 
to isolate single dot tunnelling in an impurity potential 
and quench multiple dot tunnelling. If the transverse dimension is increased too much, several 
IQD conduct for the same gate voltage ranges, causing eventually a complex structure of conductance resonances. Furthermore, we find that in the dielectric regime the variance of $lng$ systematically scales with the average of $lng$. Such an observation is consistent with a one-parameter description of the metal-insulator transition. Finally, our results strongly suggest that reducing the channel length even further should imply the observation of Coulomb oscillations with a charging energy comparable to room temperature.


\begin{thebibliography}{99}
\bibitem{scott-thomas}  J.H.F. Scott-Thomas et al. Phys. Rev. Lett. {\bf 62}, 583 (1989).

\bibitem{degraaf} C. deGraaf et al. Phys. Rev. B {\bf 44}, 9072 (1991); A. A. M. Staring et al. Phys. Rev. B {\bf 45}, 9222 (1992).
\bibitem{nicholls}  J.T. Nicholls et al. Phys. Rev. B {\bf 48}, 8866 (1993).
\bibitem{zhuang} L. Zhuang, L. Guo and S. Y. Chou, Appl. Phys. Lett. {\bf 72}, 1205 (1998). 

\bibitem{ishikuro} H. Ishikuro and T. Hiramoto, Appl. Phys. Lett. {\bf 71}, 3691 (1997).
\bibitem{kuznetsov} V. V. Kuznetsov  et al. Phys. Rev. B {\bf 56}, R15533 (1997); V. V. Kuznetsov et al. Phys. Rev. B {\bf 54}, 1502 (1996); A.K. Savchenko et al. Phys. Rev. B {\bf 52}, R17021 (1995).
\bibitem{beasley} D. Ephron et al. Phys. Rev. B {\bf 49}, 2989 (1994).
\bibitem{zorin} A. B. Zorin, Rev. Sci. Instrum. {\bf 66}, 4296 (1995).

\bibitem{threshold} The total carrier density is fixed by the threshold voltage, i.e. the gate voltage at which the inversion 2D layer is induced. It is estimated to be around zero volt for the 100 nm series ( +0.4V for the 50nm series).
\bibitem{azbel} M. Azbel Phys. Rev. B {\bf 28}, 4106 (1983).

\bibitem{fitcomment}A very small background ($< 0.1\%$ at $360mK$) has been added to the fit.

\bibitem{raikh} M. E. Raikh and I. M. Ruzin Sov. Phys. JETP {\bf 68}, 642 (1989); F. Ladieu and 
J.P. Bouchaud, J. Phys. 1 France 3, 2311 (1993). F. Bardou Europhys. Lett. {\bf 39}, 239 (1997).
\bibitem{imry}  Y. Imry Europhys. Lett. {\bf 1}, 249 (1986).
  
\bibitem{wilfrid} For too wide samples, as well as for too long wires, $var(lng) \ll - <\! lng\!>$. see
W. Poirier, D. Mailly and M. Sanquer, Phys. Rev. B {\bf 59}, 10856 (1999). The  prefactor 0.55 is smaller than expected (1) 
perhaps due to the finite temperature.
\bibitem{Pichard} J.L. Pichard and M. Sanquer in "Quantum chaos", G. Casati and
B. Chirikov editors, Cambridge University Press 1995. J.L. Pichard and M. Sanquer, Physica A {\bf 167}, 66 (1990). 

\bibitem{ruzin} I.M. Ruzin et al. Phys. Rev. B {\bf 45}, 13469 (1992).

\end{thebibliography}
\end{document}